\documentclass[preprint,superscriptaddress,prb,aps]{revtex4}

\usepackage{graphicx}

\def\BiR{Bi$_2$Sr$_{2-x}R_x$CuO$_y$}
\def\Tc{$T_{\mbox{\scriptsize c}}$}
\def\Tcmax{$T_{\mbox{\scriptsize c}}^{\mbox{\scriptsize max}}$}
\def\Tstar{$T^*$}
\def\IPG{$I_{\mbox{\scriptsize PG}}$}
\def\etal{{\it et al}.}

\begin{document}

\title{The origin of the anomalously strong influence of 
out-of-plane disorder on high-{\boldmath \Tc} superconductivity}
\author{Y. Okada}
\affiliation{Department of Crystalline Materials Science,
  Nagoya University, Nagoya 464-8603, Japan}
\author{T. Takeuchi}
\affiliation{EcoTopia Science Institute, Nagoya University, 
  Nagoya 464-8603, Japan}
\author{T. Baba}
\affiliation{Institute for Solid State Physics (ISSP), 
  University of Tokyo, Kashiwa 277-8581, Japan}
\author{S. Shin}
\affiliation{Institute for Solid State Physics (ISSP), 
  University of Tokyo, Kashiwa 277-8581, Japan}
\author{H. Ikuta}
\affiliation{Department of Crystalline Materials Science, 
  Nagoya University, Nagoya 464-8603, Japan}

\date{\hspace*{2in}}

\begin{abstract}
The electronic structure of \BiR\ ($R$=La, Eu) near 
the ($\pi$,0) point of the first Brillouin zone
was studied by means of 
angle-resolved photoemission spectroscopy (ARPES). 
The temperature \Tstar\ above which 
the pseudogap structure in the ARPES spectrum disappears 
was found to have an $R$ dependence that is opposite to that of 
the superconducting transition temperature \Tc.
This indicates that the pseudogap state is competing 
with high-\Tc\ superconductivity, and
the large \Tc\ suppression caused by out-of-plane disorder 
is due to the stabilization of the pseudogap state.
\end{abstract}

\pacs{74.25.Jb, 74.62.-c, 74.72.Hs, 74.62.Dh}
\maketitle

\begin{figure}
\includegraphics[width=0.7\columnwidth]{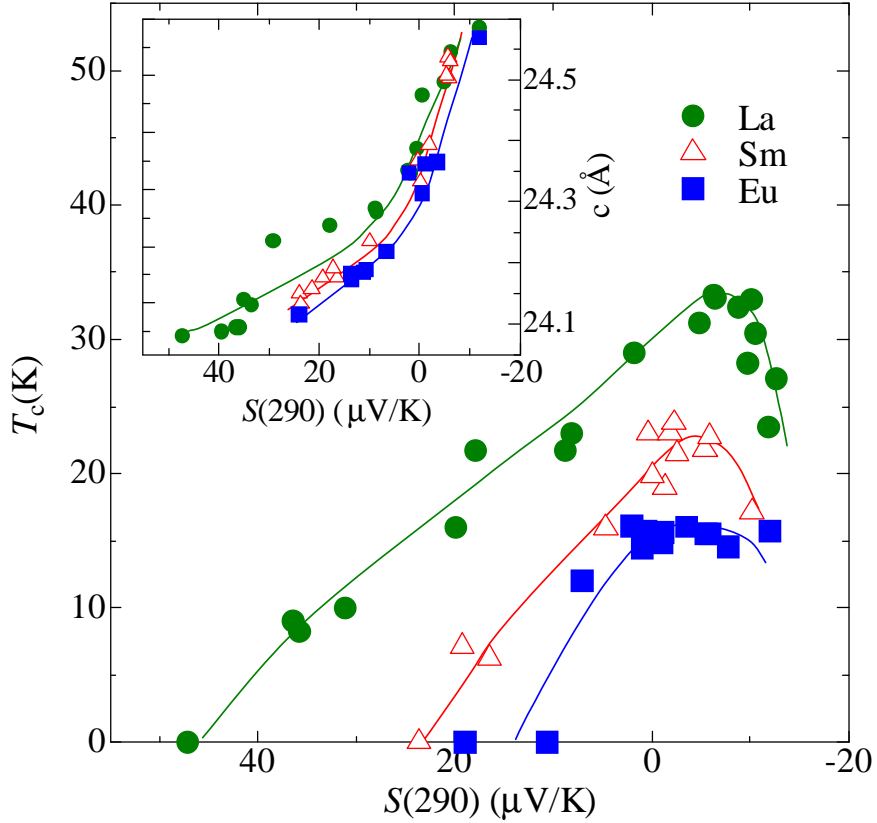}
\caption{\label{fig:TcvsS}
(color online)
The critical temperature \Tc\ as a function of $S$(290), 
the thermopower at 290 K. 
\Tc\ was determined from 
the temperature dependence of resistivity, 
which was measured simultaneously with thermopower. 
Data are based on our previous work, \cite{Okada06}
and some new data points are included. 
Inset: Lattice constant $c$ plotted as a function of $S$(290).}
\end{figure}

High temperature superconductivity occurs with doping carriers 
to a Mott insulator.
Carriers are usually doped either by varying the oxygen content 
or by an element substitution.
Unavoidably, these procedures introduce disorder that influences 
the superconducting transition temperature \Tc\ even though only sites 
outside the CuO$_2$ plane are chemically modified.
For instance, \Tc\ of the La$_2$CuO$_4$ family depends on 
the size of the cation that substitutes for La, \cite{Attfield} 
and \Tc\ of Bi$_2$Sr$_{1.6}R_{0.4}$CuO$_y$ depends 
on the $R$ element. \cite{Nameki,Eisaki}
Recently, some of the present authors have studied extensively 
the \BiR\ system using single crystals and varied the $R$ content $x$ 
over a wide range for $R$=La, Sm, and Eu. \cite{Okada06}
The results clearly show that \Tc\ at the optimal doping \Tcmax\ 
depends strongly on the $R$ element and decreases 
with the decrease in the ionic radius of $R$, 
in other words, with increasing disorder.
By plotting \Tc\ as a function of the thermopower at 290 K $S$(290), 
it was found that the range of $S$(290) values 
for samples with a non-zero \Tc\ becomes narrower 
with increasing disorder (see Fig.\ \ref{fig:TcvsS}).
Because $S$(290) correlates well with hole 
doping in many high-\Tc\ cuprates, \cite{Obertelli} 
this suggests that the doping range where superconductivity 
occurs decreases with increasing out-of-plane disorder,
in contrast to the naive expectation that 
the plot of \Tc/\Tcmax\ vs.\ doping would merge into a universal curve
for all high-\Tc\ cuprates.

Despite the strong influence on \Tc\ and on the doping range 
where superconductivity can be observed, 
out-of-plane disorder affects only weakly 
the conduction along the CuO$_2$ plane.
According to Fujita \etal, \cite{Fujita} 
out-of-plane disorder suppresses \Tc\ more than Zn 
when samples with a similar residual resistivity are compared.
This means that out-of-plane disorder influences 
\Tc\ without being a strong scatterer, 
and that this type of disorder has an unexplained effect on \Tc.
To elucidate the reason of this puzzling behavior and 
why the carrier range of high-\Tc\ superconductivity 
is affected by out-of-plane disorder, 
we studied the electronic structure of 
$R$=La and Eu crystals by means of angle-resolved 
photoemission spectroscopy (ARPES) measurements.
We particularly focused on the so-called antinodal position, 
the point where the Fermi surface crosses 
the ($\pi$,0)-($\pi$,$\pi$) zone boundary ($\bar{\mbox{M}}$-Y cut), 
due to the following reasons.
It is generally accepted that in-plane resistivity is sensitive 
to the electronic structure near 
the nodal point of the Fermi surface. \cite{Ioffe,Yoshida03,Yoshida07}
The small influence of out-of-plane disorder 
on residual resistivity hence suggests that 
the electronic structure of this region 
is not much affected, 
as Fujita \etal\ mentioned. \cite{Fujita}
Therefore, if out-of-plane disorder causes any influence 
on the electronic structure, it would be more likely to occur 
at the antinodal point of the Fermi surface.

The single crystals used in this study were grown by 
the floating zone method as reported previously. \cite{Okada06}
As mentioned in that work and commonly observed for 
Bi-based high-\Tc\ cuprates, 
the composition of the grown crystal is not 
the same as the starting one 
and depends on the position within the boule. 
Accordingly, the hole doping level can not be determined from 
the starting composition of the crystal.
On the other hand, it has been shown for many cuprates 
that $S$(290) correlates well with hole doping. 
Although $S$(290) is not directly related to the amount of carriers
and should depend on the detail of the electronic structure,
this empirical connection provides a reasonable
indicator for the hole doping level.
We note that we have confirmed in a separate experiment that 
the Fermi surface of a $R$=La and a $R$=Eu crystal 
with similar $S$(290) values coincided quite well, \cite{OkadaSNS07}
implying that their hole doping was similar.
Therefore, we use $S$(290) as a measure of doping 
in the following. \cite{NoticeThermo}

\begin{figure}
\includegraphics[width=0.9\columnwidth]{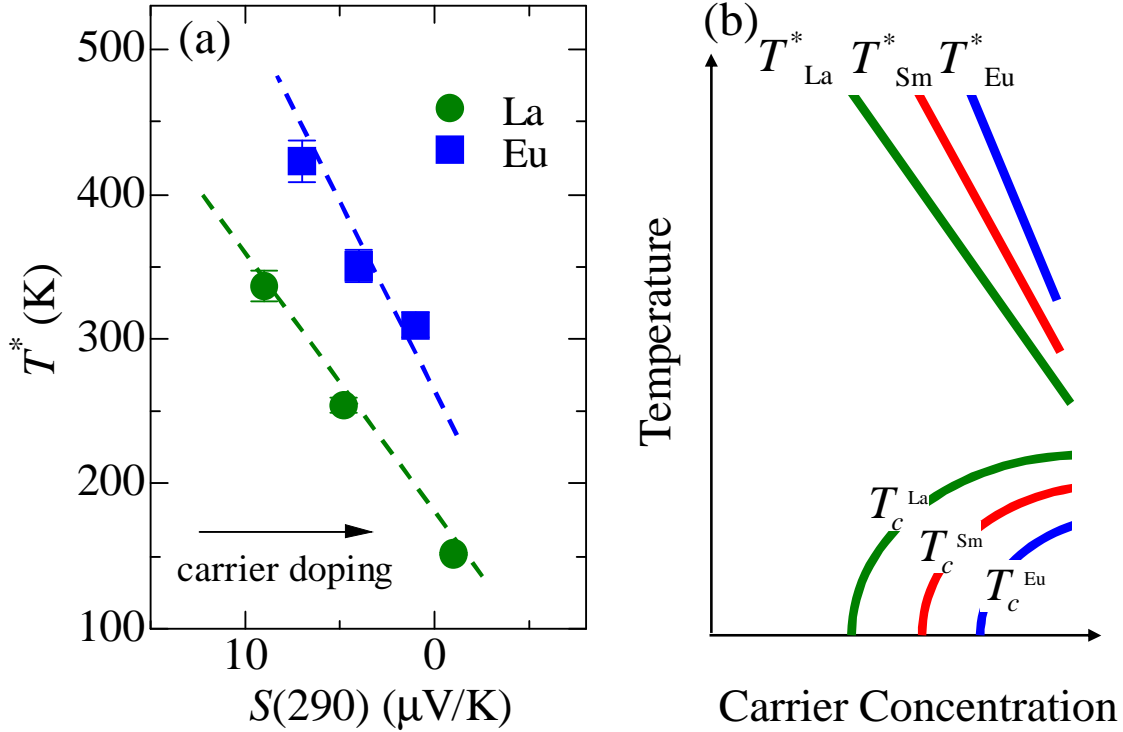}
\caption{\label{fig:diagram}
(color online)
(a) Pseudogap temperature \Tstar\ plotted as a function of $S$(290). 
(b) A schematic phase diagram of \BiR\ based on the results of 
Figs.\ \protect\ref{fig:TcvsS} and \protect\ref{fig:diagram}(a).}
\end{figure}

All crystals were annealed at 750$^\circ$C for 72 hours in air.
The ARPES spectra were accumulated using 
a Scienta SES2002 hemispherical analyzer with 
the Gammadata VUV5010 photon source (He I$\alpha$) at 
the Institute of Solid State Physics (ISSP), the University of Tokyo, 
and at beam-line BL5U of UVSOR at the Institute 
for Molecular Science, Okazaki
with an incident photon energy of 18.3 eV.
The energy resolution was 10-20 meV for all measurements, 
which was determined by the intensity reduction 
from 90\% to 10\% at the Fermi edge of a reference gold spectrum.
Thermopower was measured by a four-point method using 
a home-built equipment.
$S$(290) was determined using crystals
that were cleaved from those used for ARPES measurements 
except the $R$=La sample that had the largest doping in 
Fig.\ \ref{fig:diagram}(a). 
For that particular sample, the $S$(290) value was estimated 
from the $c$-axis length deduced from x-ray diffraction based
on the data shown in the inset to Fig.\ \ref{fig:TcvsS}. 

\begin{figure}
\includegraphics[width=0.98\columnwidth]{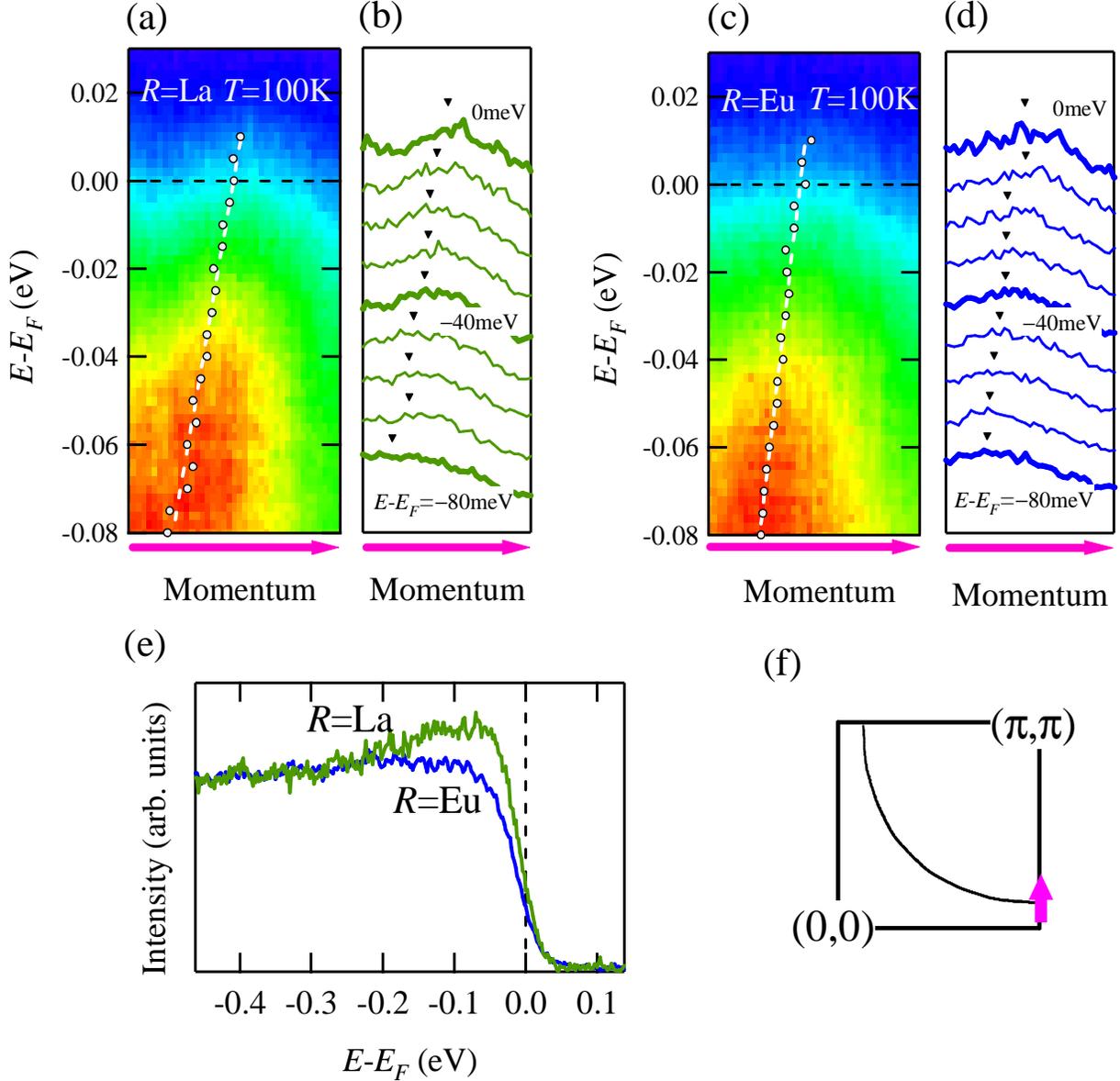}
\caption{\label{fig:IntPlot}
(color online)
Intensity plots in the energy-momentum plane of 
the ARPES spectra at 100 K of slightly underdoped \BiR\ samples 
that have a similar doping level with 
(a) $R$=La and (c) $R$=Eu along the momentum line indicated by
the arrow in (f).
(b), (d) Momentum distribution curves (MDCs) of the two samples. 
(e) The energy distribution curves (EDCs) of the two samples at $k_F$. 
(f) Schematic drawing of the underlying Fermi surface.}
\end{figure}

Figures \ref{fig:IntPlot}(a) and (c) show 
the ARPES intensity plots along 
the ($\pi$,0)-($\pi$,$\pi$) direction 
at 100 K for $R$=La and Eu crystals that have 
a similar hole concentration.
The samples were cleaved in situ at 250 K 
in a vacuum of better than 5$\times$10$^{-11}$ Torr.
The $S$(290) values were 4.7 $\mu$V/K and 4.8 $\mu$V/K for 
the $R$=La and Eu samples, respectively, 
indicating that they are slightly underdoped 
(see Fig.\ \ref{fig:TcvsS}).
Figures \ref{fig:IntPlot}(b) and (d) show momentum 
distribution curves (MDCs) of the $R$=La and Eu samples, respectively.
We fitted the MDC curves to a Lorentz function to determine the peak position.
The thus extracted dispersion is superimposed by white small circles 
on Figs.\ \ref{fig:IntPlot}(a) and (c).
The momentum where the dispersion curve crosses 
the Fermi energy $E_F$ corresponds to the Fermi wave vector $k_F$ 
on the ($\pi$,0)-($\pi$,$\pi$) cut, 
and Fig.\ \ref{fig:IntPlot}(e) shows the energy distribution 
curves (EDCs) of the two samples at $k_F$.
Obviously, the $R$=La sample has 
a larger spectral weight at $E_F$,
although the doping level of the two samples is very similar. 

\begin{figure}
\includegraphics[width=0.85\columnwidth]{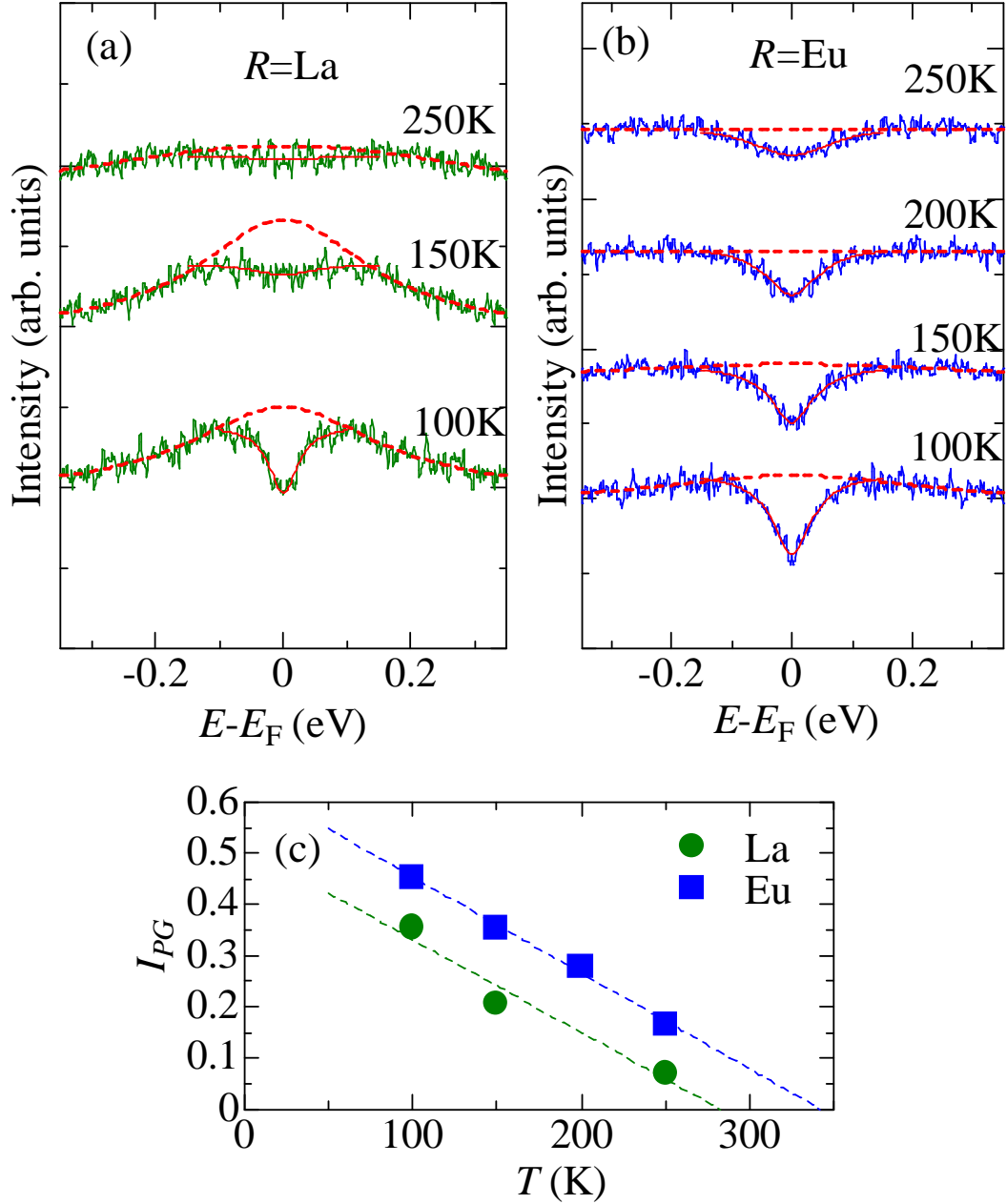}
\caption{\label{fig:EDC}
(color online)
Temperature dependence of the symmetrized ARPES spectrum at 
the antinodal point for the (a) $R$=La and (b) $R$=Eu crystals 
of Fig.\ \protect\ref{fig:IntPlot}. 
(c) Temperature dependence of the amount 
of spectral weight suppression \IPG.}
\end{figure}

Figure \ref{fig:EDC} shows the EDCs of 
the two samples of Fig.\ \ref{fig:IntPlot} at various temperatures. 
To remove the effects of the Fermi function on 
the spectra, we applied the symmetrization method 
$I_{\mbox{\scriptsize sym}}(\omega)=I(\omega)+I(-\omega)$, 
where $\omega$ denotes the energy relative to $E_F$. \cite{Norman98} 
As shown in Figs.\ \ref{fig:EDC}(a) and (b), 
the symmetrized spectra of both samples show clearly 
a gap structure at the lowest measured temperature, 100 K. 
Because we are probing the antinodal direction at 
a temperature that is higher than \Tc, 
we attribute this gap structure to the pseudogap. 
With increasing the temperature, the gap structure fills up 
without an obvious change in the gap size.
At 250 K, only a small suppression of the spectral weight 
was observed for the $R$=La sample. 
On the other hand, a clear pseudogap structure 
can be observed for the $R$=Eu sample even at 250 K. 
This means that the temperature \Tstar\ up to which 
the pseudogap structure can be observed is certainly different 
despite the closeness of the doping level. 

The thin solid lines $I_{\mbox{\scriptsize fit}}(\omega)$ of 
Figs.\ \ref{fig:EDC}(a) and (b) are the results of 
fitting a Lorentz function to the symmetrized spectrum 
in the energy range of $E_F\pm150$ meV. 
The dashed lines are, on the other hand, the background spectra 
$I_{\mbox{\scriptsize bkg}}(\omega)$, 
which were determined by making a similar fit 
in the energy range of 150 meV$\leq|\omega|\leq$400 meV. 
It can be seen that the difference between the two fitted curves 
at $E_F$ increases with decreasing the temperature, 
reflecting the growth of the pseudogap. 
To quantify how much the spectral weight at $E_F$ is depressed, 
we define \IPG\ as the difference between unity and 
the spectral weight of the fitted spectrum at $E_F$ divided 
by that of the background curve 
($1-I_{\mbox{\scriptsize fit}}(0)/I_{\mbox{\scriptsize bkg}}(0)$). 
Figure \ref{fig:EDC}(c) shows the temperature dependence 
of \IPG\ of the two samples. 
Obviously, the depression of the spectral weight is larger for 
the $R$=Eu sample at all measured temperatures, and \Tstar\ is higher. 
\IPG\ is roughly linear temperature dependent 
for both samples with a very similar slope. 
Therefore, we extrapolated the data 
with the same slope as shown by the dashed lines, 
and estimated \Tstar\ to be 282 K and 341 K for 
the $R$=La and Eu samples, respectively. 

We also measured ARPES spectra of four other samples at 150 K. 
Assuming that the temperature dependence of \IPG\ is the same 
as that of Fig.\ \ref{fig:EDC}(c), 
we can estimate \Tstar\ from the \IPG\ value at 150 K. 
The thus estimated \Tstar\ values are included in Fig.\ \ref{fig:diagram}(a), 
which shows \Tstar\ of all samples studied in this work 
as a function of $S$(290). 
It can be seen that \Tstar\ is higher for $R$=Eu than $R$=La 
when compared at the same $S$(290) value. 
Because \Tc\ at the same hole doping decreases with changing 
the $R$ element to one with a smaller ionic radius (Fig.\ \ref{fig:TcvsS}), 
it is clear that \Tc\ and \Tstar\ have 
an opposite $R$ dependence. 
This important finding is summarized on the schematic phase diagram 
shown in Fig.\ \ref{fig:diagram}(b). 
As shown, both \Tcmax\ and the carrier range where superconductivity 
takes place on the phase diagram decrease with decreasing 
the ionic radius of $R$, 
while \Tstar\ at the same hole concentration 
increases. \cite{NoticeThermo}

One of the most important issues of pseudogap has been 
whether such state is a competitive one 
or a precursor state of high-\Tc\ superconductivity. \cite{Norman05} 
Figure \ref{fig:diagram}(b) clearly shows that whatever 
the microscopic origin of the pseudogap is, 
it is competing with high-\Tc\ superconductivity. 
In contrast, some other studies have concluded that 
pseudogap is closely related to 
the superconducting state because the momentum dependence of the gap is 
the same above and below \Tc\ and the evolution of 
the gap structure through \Tc\ is smooth. \cite{Ding,Renner,Norman98} 
We point out, however, that several recent experiments revealed 
the existing of two energy gaps at a temperature 
well below \Tc\ for underdoped cuprate 
superconductors \cite{Tacon,Tanaka} 
as well as for optimally doped and overdoped 
(Bi,Pb)$_2$(Sr,La)$_2$CuO$_{6+\delta}$. \cite{Kondo,Boyer} 
The energy gap observed in the antinodal region was attributed 
to the pseudogap while the one near the nodal direction 
to the superconducting gap. 
We think that the conflict encountered in the pseudogap issue 
arose because distinguishing these 
two gaps would be not easy when their magnitudes are similar. 

We next discuss why \Tc\ decreases when \Tstar\ increases.
We think that the ungapped portion of the Fermi surface 
above $T_c$ is smaller 
for $R$=Eu when compared at the same doping 
and at the same temperature because pseudogap opens 
at a temperature that is higher than $R$=La. 
Indeed, the results of our ARPES experiment on optimally doped 
\BiR\ confirmed this assumption
by demonstrating that the momentum region where a quasiparticle
or a coherence peak was observed was narrower for 
the $R$=Eu sample than the $R$=La sample. \cite{OkadaSNS07}
In other words, the Fermi arc shrinks with changing 
the $R$ element from La to Eu, which mimics the behavior 
observed when doping is decreased. \cite{Yoshida03,Shen}
Because the superfluid density $n_s$ decreases 
with underdoping, \cite{Uemura} 
it is reasonable to assume that only the states on 
the Fermi arc can participate to superconductivity. 
If we can think in analogy to the carrier 
underdoping case therefore, 
$n_s$ would be smaller for $R$=Eu than $R$=La, 
and the decrease of \Tc\ is quite naturally explained from 
the Uemura relation. \cite{Uemura} 
The faster disappearance of superconductivity 
with carrier underdoping for $R$=Eu (see Fig.\ \ref{fig:TcvsS}) 
is also a straightforward consequence of this model. 
Furthermore, the opening of the pseudogap at the antinodal direction 
would not increase much the residual in-plane resistivity because 
the in-plane conduction is mainly governed by 
the nodal carriers. \cite{Ioffe,Yoshida03,Yoshida07} 
Hence, the observation that out-of-plane disorder 
largely suppresses \Tc\ with only a slight increase in 
residual resistivity \cite{Fujita} can also be 
immediately understood. 

Finally, we discuss our results in conjunction with 
the reported data of 
scanning tunneling microscopy/spectroscopy (STM/STS) experiments,
which unveiled a strong inhomogeneity in the local electronic structure 
for Bi$_2$Sr$_2$CaCu$_2$O$_y$. \cite{Howald,Pan,Lang} 
It was demonstrated that the volume fraction of the pseudogapped region 
increases with underdoping. 
The ARPES experiments, on the other hand, 
show that the spectral weight at the chemical potential of 
the antinodal region decreases with carrier underdoping. \cite{Norman05} 
Hence, the antinodal spectral weight at $E_F$ is likely 
to correlate with the volume fraction of the pseudogap region. 
In the present work, we increased the degree of out-of-plane disorder
while the hole doping was unaltered,
and observed that the spectral weight at the chemical potential
is lower for the $R$=Eu sample when compared at the same temperature,
as shown in Fig.\ \ref{fig:EDC}(c). 
We thus expect that the fraction of superconducting region 
in real space is smaller for $R$=Eu. 
Indeed, quite recent STM/STS experiments on optimally 
doped \BiR\ report that the averaged gap size is larger 
when the ionic radius of $R$ is smaller, \cite{Sugimoto}
which is attributable to an increase of the pseudogapped region. 
Moreover, our results complement the STM/STS data 
and indicate that not only the area where 
a pseudogap is observed at low temperature 
but also \Tstar\ increases with disorder
which means that the pseudogap state is stabilized and persists
up to higher temperatures.
Further, while the STM/STS studies on \BiR\ investigated only 
optimally doped samples, \cite{Sugimoto,Machida} 
we have varied hole doping,
which further corroborates the conclusion that 
the pseudogap state competes with high-\Tc\ superconductivity.

In summary, we have studied the mechanism why \Tcmax\ and 
the carrier range where high-\Tc\ superconductivity occurs 
strongly depend on the $R$ element in the \BiR\ system 
by investigating the electronic structure 
at the antinodal direction of the Fermi surface of $R$=La and Eu samples. 
We observed a pseudogap structure in the ARPES spectrum 
up to a higher temperature for $R$=Eu samples 
when samples with a similar hole doping are compared, 
which clearly indicates that the pseudogap state 
is competing with high-\Tc\ superconductivity. 
This result suggests that out-of-plane disorder increases 
the pseudogapped region and reduces the superconducting fluid density, 
which explains its strong influence on high-\Tc\ superconductivity. 
We stress that the present results are relevant 
to all high-\Tc\ superconductors 
because they are more or less suffered from out-of-plane disorders.

We would like to thank T.\ Ito of UVSOR and 
T.\ Kitao and H.\ Kaga of Nagoya University for experimental assistance.


\begin{thebibliography}{99}
\bibitem{Attfield}J. P. Attfield, A. L. Kharlanov, 
  and J. A. McAllister, 
  Nature {\bf 394}, 157 (1998).
\bibitem{Nameki}H. Nameki, M. Kikuchi, and Y. Syono, 
  Physica C {\bf 234}, 255 (1994).
\bibitem{Eisaki}H. Eisaki, N. Kaneko, D. L. Feng, A. Damascelli, 
  P. K. Mang, K. M. Shen, Z.-X. Shen, and M. Greven, 
  Phys. Rev. B {\bf 69}, 064512 (2004).
\bibitem{Okada06}Y. Okada and H. Ikuta, 
  Physica C {\bf 445-448}, 84 (2006).
\bibitem{Obertelli}S. D. Obertelli, J. R. Cooper, and J. L. Tallon, 
  Phys. Rev. B {\bf 46}, 14928 (1992).
\bibitem{Fujita}K. Fujita, T. Noda, K. M. Kojima, H. Eisaki, 
  and S. Uchida, 
  Phys. Rev. Lett. {\bf 95}, 097006 (2005).
\bibitem{Ioffe}L. B. Ioffe and A. J. Millis, 
  Phys. Rev. B {\bf 58}, 11631 (1998). 
\bibitem{Yoshida03}T. Yoshida, X. J. Zhou, T. Sasagawa, W. L. Yang, 
  P. V. Bogdanov, A. Lanzara, Z. Hussain, T. Mizokawa, A. Fujimori, 
  H. Eisaki, Z.-X. Shen, T. Kakeshita, and S. Uchida, 
  Phys. Rev. Lett. {\bf 91}, 027001 (2003).
\bibitem{Yoshida07}T. Yoshida, X. J. Zhou, D. H. Lu, S. Komiya, 
  Y. Ando, H. Eisaki, T. Kakeshita, S. Uchida, Z. Hussain, 
  Z. X. Shen, and A. Fujimori, 
  J. Phys.: Condens. Matter {\bf 19}, 125209 (2007).
\bibitem{OkadaSNS07}Y. Okada, T. Takeuchi, A. Shimoyamada, S. Shin, 
  and H. Ikuta,
  J. Phys. Chem. Solids ({\it in press}) (cond-mat/0709.0220).
\bibitem{NoticeThermo}
  There remains a chance that the hole doping of 
  La- and Eu-doped samples with the same $S$(290) value 
  is slightly different
  especially when the $R$ content increases with underdoping
  because it was reported that disorder increased thermopower 
  of La$_2$CuO$_4$-based superconductors. 
  (J. A. McAllister and J. P. Attfield, 
  Phys. Rev. Lett. {\bf 83}, 3289 (1999).)
  However, this will not affect our conclusion, 
  because if this is the case, the doping of a disordered
  sample would be larger than we are assuming
  and the data of the Eu-doped samples of 
  Fig.\ \protect\ref{fig:diagram}(b)
  shift slightly to the more carrier doped side 
  relative to the La-doped samples.
  This is in favor to our conclusion.
\bibitem{Norman98}M. R. Norman, H. Ding, M. Randeria, J. C. Campuzano, 
  T. Yokoya, T. Takeuchi, T. Takahashi, T. Mochiku, K. Kadowaki, 
  P. Guptasarma, and D. G. Hinks, Nature {\bf 392}, 157 (1998).
\bibitem{Shen}K. M. Shen, F. Ronning, D. H. Lu, F. Baumberger, 
  N. J. C. Ingle, W. S. Lee, W. Meevasana, Y. Kohsaka, M. Azuma, 
  M. Takano, H. Takagi, and Z.-X. Shen, Science {\bf 307}, 901 (2005).
\bibitem{Norman05}M. R. Norman, D. Pines, and C. Kallin, 
  Adv. Phys. {\bf 54}, 715 (2005).
\bibitem{Ding}H. Ding, T. Yokoya, J. C. Campuzano, T. Takahashi, 
  M. Randeria, M. R. Norman, T. Mochiku, K. Kadowaki, 
  and J. Giapintzakis, 
  Nature {\bf 382}, 51 (1996).
\bibitem{Renner}Ch. Renner, B. Revaz, J. Y. Genoud, K. Kadowaki, 
  and \O. Fischer, 
  Phys. Rev. Lett. {\bf 80}, 149 (1998).
\bibitem{Tacon}M. Le Tacon, A. Sacuto, A. Georges, G. Kotliar, 
  Y. Gallais, D. Colson, and A. Forget, 
  Nature Physics {\bf 2}, 537 (2006).
\bibitem{Tanaka}K. Tanaka, W. S. Lee, D. H. Lu, A. Fujimori, T. Fujii, 
  Risdiana, I. Terasaki, D. J. Scalapino, T. P. Devereaux, Z. Hussain, 
  and Z.-X. Shen, 
  Science {\bf 314}, 1910 (2006).
\bibitem{Kondo}T. Kondo, T. Takeuchi, A. Kaminski, S. Tsuda, 
  and S. Shin, 
  Phys. Rev. Lett. {\bf 98}, 267004 (2007).
\bibitem{Boyer}M. C. Boyer, W. D. Wise, K. Chatterjee, M. Yi, T. Kondo, 
  T. Takeuchi, H. Ikuta, and E. W. Hudson, 
  Nature Phys. {\bf 3}, 802 (2007).
\bibitem{Uemura} Y. J. Uemura, G. M. Luke, B. J. Sternlieb, 
  J. H. Brewer, J. F. Carolan, W. N. Hardy, R. Kadono, 
  J. R. Kempton, R. F. Kief, S. R. Kreitzman, P. Mulhern, 
  T. M. Riseman, D. L. Williams, B. X. Yang, S. Uchida, 
  H. Takagi, J. Gopalakrishnan, A. W. Sleight, M. A. Subramanian, 
  C. L. Chien, M. Z. Cieplak, Gang Xiao, V. Y. Lee, B. W. Statt, 
  C. E. Stronach, W. J. Kossler, and X. H. Yu, 
  Phys. Rev. Lett. {\bf 62}, 2317 (1989).
\bibitem{Howald}C. Howald, P. Fournier, and A. Kapitulnik, 
  Phys. Rev. B {\bf 64}, 100504(R) (2001).
\bibitem{Pan}S. H. Pan, J. P. O'Neal, R. L. Badzey. C. Chamon, 
  H. Ding, J. R. Engelbrecht, Z. Wang, H. Eisaki, S. Uchida, 
  A. K. Guptak, K. W. Ng, E. W. Hudson, K. M. Lang, 
  and J. C. Davis, 
  Nature {\bf 413}, 282 (2001).
\bibitem{Lang}K. M. Lang, V. Madhavan, J. E. Hoffman, E. W. Hudson, 
  H. Eisaki, S. Uchida, and J. C. Davis, 
  Nature {\bf 415}, 412 (2002).
\bibitem{Sugimoto}A. Sugimoto, S. Kashiwaya, H. Eisaki, H. Kashiwaya, 
  H. Tsuchiura, Y. Tanaka, K. Fujita, and S. Uchida, 
  Phys. Rev. B {\bf 74}, 094503 (2006).
\bibitem{Machida}T. Machida, Y. Kamijo, K. Harada, T. Noguchi,
  R. Saito, T. Kato, and H. Sakata,
  J. Phys. Soc. Jpn. {\bf 75}, 083708 (2006).
\end{thebibliography}
\end{document}